\documentclass[lettersize,journal]{IEEEtran}

\IEEEoverridecommandlockouts
\usepackage{amsmath,amsfonts}
\usepackage{algorithm}
\usepackage{array}
\usepackage{bbm}
\usepackage[caption=false,font=normalsize,labelfont=sf,textfont=sf]{subfig}
\usepackage{textcomp}
\usepackage{stfloats}
\usepackage{url}
\usepackage{verbatim}
\usepackage{cite}
\usepackage{booktabs}
\usepackage{tabularx}
\usepackage{amsthm}
\usepackage{balance}
\usepackage[bookmarks=false]{hyperref}
\usepackage{footmisc}
\usepackage[english]{babel}
\usepackage{graphicx}
\usepackage{cleveref}
\usepackage{dsfont}
\usepackage{algpseudocode}
\renewenvironment{thebibliography}[1]{
  \begin{oldthebibliography}{#1}
    \setlength{\itemsep}{0pt}
    \setlength{\parsep}{0pt}
}{
  \end{oldthebibliography}
}
\DeclareMathOperator*{\argmax}{argmax}

\theoremstyle{definition}
\newtheorem{theorem}{Theorem}
\newtheorem{proposition}{Proposition}

\theoremstyle{remark}

\begin{document}

\title{Information Value-Based Communications for Remote Inference over Wireless Networks}

\author{
    \IEEEauthorblockN{Nitisha Singh\IEEEauthorrefmark{1}, Walid Saad\IEEEauthorrefmark{1}, and Pascal Adjakpl\'e\IEEEauthorrefmark{2}\author{
    \IEEEauthorblockN{Nitisha Singh\IEEEauthorrefmark{1}, Walid Saad\IEEEauthorrefmark{1}, and Pascal Adjakpl\'e\IEEEauthorrefmark{2}\thanks{This work was supported by InterDigital.}} \\
    \IEEEauthorblockA{\small \IEEEauthorrefmark{1}Bradley Department of Electrical and Computer Engineering, Virginia Tech, Alexandria, VA, 22305, USA} \\
    \IEEEauthorblockA{\small \IEEEauthorrefmark{2}InterDigital, USA} \\
    \IEEEauthorblockA{ Emails: \{nitishas, walids\}@vt.edu, pascal.adjakple@interdigital.com\vspace{-10mm}}
}} \\
    \IEEEauthorblockA{\small \IEEEauthorrefmark{1}Bradley Department of Electrical and Computer Engineering, Virginia Tech, Alexandria, VA, 22305, USA} \\
    \IEEEauthorblockA{\small \IEEEauthorrefmark{2}InterDigital, USA} \\
    \IEEEauthorblockA{ Emails: \{nitishas, walids\}@vt.edu, pascal.adjakple@interdigital.com\vspace{-10mm}}
}



\maketitle
\thispagestyle{empty}

\begin{abstract}
Physical artificial intelligence (AI) is expected to be an important use case for emerging 6G networks, with large AI-based inference models requiring communication from distributed edge devices. To support remote inference tasks for physical AI, one can leverage the concept of task-oriented communications, which involves transmitting only task-relevant features to improve communication efficiency and remote inference performance. However, the prior art in task-oriented communication systems does not account for the information value (IV) of the data for network management and instead rely on age or utility-based proxies. In this paper, a novel IV-aware resource allocation framework is proposed for a physical remote inference system to improve downstream task performance by prioritizing the transmission of high-IV information. The Shapley value concept is then used to formalize the IV notion for quantifying the marginal contribution of individual data streams to the overall inference task. In order to compute this value in practical systems, a novel online algorithm is proposed to track the possibly dynamic IV with $O(N)$ complexity. Then, the IV-aware resource block allocation problem is posed as a constrained Markov decision process and reduced to a deterministic per-slot problem using the Lyapunov drift plus penalty (LDPP) approach. Experiments for the 3D object detection task using the BEVFusion model on nuScenes-mini dataset show that the proposed IV-LDPP scheduler achieved a 53\% higher task performance than the value of information update (VoIU) baseline while using 43\% less bandwidth than the MaxRate baseline across varying channel conditions. 
\end{abstract}
\vspace{-2mm}
\begin{IEEEkeywords}
Information Value, Remote Inference, Task-Oriented Communication, Shapley Value, Physical AI.
\end{IEEEkeywords}
\vspace{-6mm}
\section{Introduction}
\vspace{-1mm}
Next-generation wireless systems, such as 6G and beyond, must support emerging services such as cyber-physical systems, extended reality (XR), and physical artificial intelligence (AI) \cite{6gusecases}. Central to these is the task of remote inference in which distributed edge devices (e.g. sensors, drones, or robots) continuously offload high-dimensional sensor data to edge servers equipped with advanced AI models\cite{6gedgeai}. Since AI-based inference models work with specific latent structure and patterns rather than requiring perfect data reconstruction, a perfect bit-by-bit transmission of the payload is unnecessary and could waste the resources of the wireless network. Instead, a promising approach to improve communication efficiency is to extract and transmit only the task-relevant features \cite{shao2023learningtaskorientedcommunicationedge}. For instance, in multi-view 3D object detection, transmitting background pixels that lack dynamic entities could end up using critical resources that should be instead allocated to agents capturing target-dense regions of interest. This requires a shift from traditional bit transmission to \emph{task-oriented (or goal-oriented) communications} driven by the information value (IV) of the data \cite{semcomdata}. In the context of remote inference, IV represents the meaning, utility, and relevance of a specific data sample to the accuracy of the downstream for the served application. By characterizing the IV, the network can intelligently prioritize the transmission of only the most critical task-relevant features. As such, an IV-driven approach can effectively conserve scarce uplink resources while maximizing the expected inference gain. 
\par In fact, the design of IV-driven networks serves as a precursor to the broader realization of semantic communications (SC) \cite{lesdata}. However, translating IV into a practical scheduling policy is challenging. Data-valuation metrics in the field of machine learning (ML) are typically evaluated post-inference and are tied to the specific model (particularly to the inference application) \cite{ghorbani2019data}. This is not applicable in wireless resource allocation because the scheduler must quantify the IV a priori, before allocating resources, and therefore, before the inference task is actually executed. Moreover, even if these post-inference data valuation metrics can be adapted, calculating them requires extensive computational overhead (e.g. combinatorial evaluations), which makes them impractical for real-time scheduling. Therefore, to develop an IV-aware scheduler for supporting the remote inference needs of physical AI, we must first rigorously define the IV notion and, then, leverage it for efficient resource allocation.
\vspace{-4mm}
\subsection{Related Works}
\vspace{-1mm}
The current literature on resource allocation for remote inference has heavily relied on age of information (AoI) to design scheduling policies under bandwidth constraints \cite{10559951, 11206222, zhao2022aoi}. Several other works have adapted the traditional value of information (VoI) for specific system goals, such as control and vehicle-to-everything (V2X) communications \cite{valuematters, lei2025learningvalueinformationjoint}. While these metrics are typically used as proxies for data utility, they fail to capture the actual value for supporting remote inference tasks for physical AI use cases. On the one hand, AoI-based schemes optimize for data freshness, meaning the scheduler would prioritize a stale user, even if its information is redundant for inference. On the other hand, existing VoI frameworks define information value as the reduction in quadratic control cost or tracking error. While well-suited for dynamic control objectives, this definition requires an analytical model that can directly relate the transmitted observation to the performance metric, which is not always available in AI inference models.
\par Unlike the literature on AoI and VoI, the literature in goal-oriented communication systems has focused on task-relevant source encoding \cite{attnsemcolinf, llcinfsem}. 
In \cite{attnsemcolinf}, a transformer-based SC system is proposed to select only the task-relevant patches for image transmission. In \cite{llcinfsem}, a multi-stage encoder is developed for optimizing latency in multi-state edge inference. The approaches of \cite{attnsemcolinf} and \cite{llcinfsem} optimize per-user task-relevant feature encoding, but they do not provide a mechanism to quantify the relative importance of the users to the inference task via an IV metric, which is necessary for value-aware resource allocation.
\par Here, we note that, in \cite{semcomdata}, the authors highlighted the need for a semantics-based metric for random access in SC systems. However,  the discussion in \cite{semcomdata} is limited to freshness and limits its definition of value to a cost vs benefit tradeoff. The authors in \cite{10494374} proposed a metric for semantic value based on the Zipf distribution; however, its application is limited to text transmission. Thus, the existing literature on goal-oriented SC and remote inference in \cite{10559951, 11206222, zhao2022aoi, valuematters, lei2025learningvalueinformationjoint, attnsemcolinf, llcinfsem, semcomdata, 10494374} does not provide an application-agnostic IV metric that extends beyond AoI and VoI metrics to quantify the dynamic value for inference applications. Without such an IV metric, the network cannot identify the most important users for achieving the optimal inference performance and would thus waste valuable network resources.\vspace{-21pt}
\subsection{Contributions}
\vspace{-2mm}
The main contribution of this paper is a novel framework to quantify IV and to leverage it for optimal resource allocation to improve inference performance for supporting the remote inference tasks of physical AI agents (e.g. robots), when they are deployed over a resource-constrained wireless network. In particular, we consider a wireless network that must support, in the uplink, a group of distributed robots engaged in a collaborative remote inference task (e.g. object detection, cooperative semantic segmentation). We propose an IV-aware scheduler that allows the network to utilize IV at the base station for resource allocation to capture the importance of a robot's features relative to the features obtained from other robots, via the concept of Shapley value \cite{williamson2020efficientnonparametricstatisticalinference}.
Since computing the Shapley value incurs significant computational overhead, we propose an online stochastic gradient descent (SGD)-based algorithm to handle non-stationary environmental data distributions and evolving feature importance. This algorithm enables real-time tracking of the global IV with $O(N)$ complexity by avoiding the computational cost of periodically estimating  IV or using outdated estimates. We formulate the IV-aware resource allocation problem in a way to maximize the IV-weighted transmission success while constraining the long-term latency as a constrained Markov decision (CMDP). Then, we propose a solution based on Lyapunov drift plus penalty (LDPP). The IV-LDPP resource allocation scheme is compared across conventional capacity-driven (MaxRate) and freshness-driven (AoI) as well as task-aware baselines such as task-aware age of information (TAoI), value of information update (VoIU) for an object detection task using the birds-eye view (BEV) fusion model using the nuScenes mini dataset. The proposed IV-LDPP scheduler achieves a 53\% higher task performance than the VoIU baseline while using 43\% less bandwidth than the MaxRate baseline across varying channel conditions. 

\vspace{-4mm}
\section{System Model}
\vspace{-1mm}
\begin{figure}[t]
    \centering
    \includegraphics[width = \linewidth]{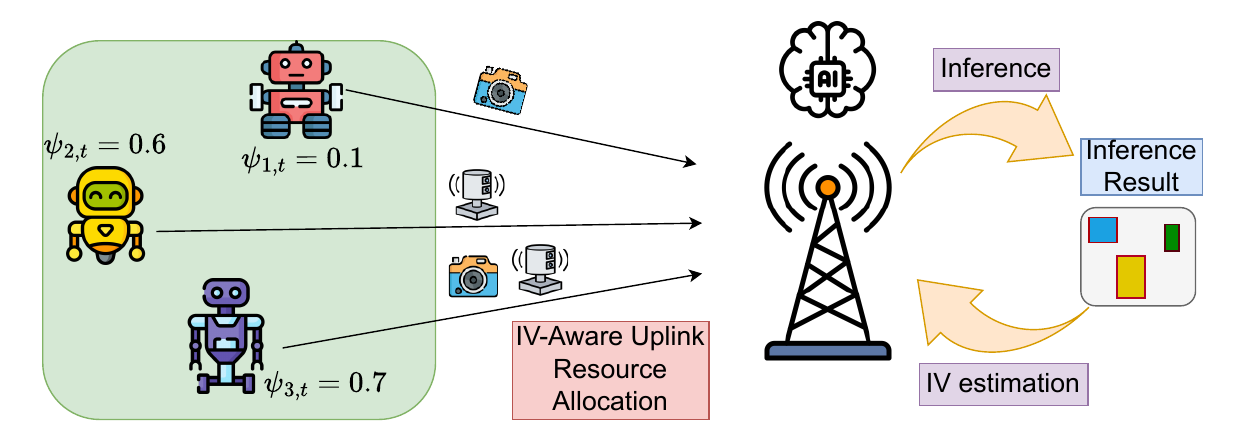}
    \caption{Illustration of the proposed IV-aware resource allocation framework for physical AI. Robots with different modality capabilities are engaged in the 3D object detection task. To overcome bandwidth constraints, the BS allocates limited uplink RBs based on the IV metric $\psi_{t}$, to maximize communication efficiency and inference performance.}
    \label{systemmodel}
    \vspace{-3mm}
\end{figure}
We consider a single-cell goal-oriented wireless network involving a set of robots $\mathcal{N}$ and a base station (BS) operating in a time-indexed manner $t\in \{0, 1,... \}$. Each robot $n \in \mathcal{N}$ acts as an independent physical AI equipped with diverse sensors (e.g. LiDAR, radar, RGB camera). At each time step $t$,  robot $n$ captures a raw, high-dimensional observation $\boldsymbol{x}_n$ of its local environment and transmits it to the BS for remote inference, as shown in Fig. \ref{systemmodel}. A goal-oriented remote inference application, represented by a function $f$, is executed at the BS, requiring observations from the robots to infer a target variable $y$. To facilitate this task, the robots must transmit their observations over the uplink to the BS, which serves as a central data sink. However, we assume that the uplink bandwidth is constrained and divided into a finite set of orthogonal frequency division multiple access (OFDMA) resource blocks (RBs). Since the raw observation $\boldsymbol{x}_n$ is high-dimensional, transmitting it directly under constrained bandwidth is sub-optimal for inference latency. Therefore, the BS must schedule limited resource units only to the subset of robots whose data provides the highest utility for inferring $y$.
\vspace{-4mm}
\subsection{Communication Model}
\vspace{-1mm}
At each time slot $t$, each robot $n \in \mathcal{N}$ generates a high-dimensional observation vector, $\boldsymbol{x}_{n,t}\in \mathbb{R}^{D_n}$, by sampling its dynamic local environment. To overcome limited uplink communication bandwidth, each robot extracts task-relevant features to compress its observations $\boldsymbol{x}_{n,t}$ into a low-dimensional representation $\boldsymbol{z}_{n,t} \in \mathbb{R}^{d_n}$, where $d_n << D_n$ to preserve only the task-relevant features in the observation.
 \par The communication between the robots and the BS occurs over a shared wireless channel. The BS controls the uplink wireless channel, which is divided into $M$ RBs, in each time slot $t \in \{0, 1,..\}$. The BS's scheduling decision at time $t$ is represented by $r_{n,t}$, which captures the number of RBs allocated to robot $n$ at time $t$.
Let $B$ be the bandwidth of a single RB. The quality of the wireless link between robot $n$ and the BS is time-varying and a function of its distance $d_{n,t}$ from the BS at time $t$, as the robot navigates its operational area. The considered channel consists of large-scale path loss and small-scale fading \cite{10499876}, and the complex baseband channel coefficient is given by, $h_{n,t} = \sqrt{l(d_{n,t})}\tilde{h}_{n,t}$,
where $l(d_{n,t}) = d_{n,t}^{-\alpha}$ represents the large-scale path loss with exponent $\alpha$ and $\tilde{h}_{n,t} \sim \mathcal{N}(0,1)$ models the Rayleigh small-scaled fading. Consequently, the channel power gain $|h_{n,t}|^2$ is exponentially distributed with mean $l(d_{n,t})$.
 Since the BS allocates orthogonal RBs to each robot, there is no intra-cell interference. However, we account for constant inter-cell interference from neighboring deployments. The instantaneous signal-to-interference-plus-noise ratio (SINR) of the channel between robot $n$ and the BS at a time $t$ can be given by, $ \gamma_{n,t} = \frac{P |h_{n,t}|^2}{\sigma_{\textrm{noise}}^2 + I_{\textrm{inter}}}$ where $P$ is the constant transmit power, $\sigma^{2}_{\text{noise}}$ is the noise power, and $I_{\text{inter}}$ is the average out of cell interference power. 
When robot $n$ is allocated resources at time $t$, the probability of successful transmission depends upon the likelihood of the distortion and latency of the transmission being below a certain threshold. Let $\epsilon_n$ be the maximum tolerable distortion and $D_{\textrm{max}}$ be the maximum allowable end-to-end delay for robot $n$, then the reliability of the transmission is given as,
Here, $T_n$ is the transmission delay, which is defined as,
\vspace{-2mm}
\begin{equation}
    T_n(r_{n,t},\gamma_{n,t}) = \frac{L_n}{r_{n,t} B_m \log_2 (1 + \gamma_{n,t})},
\end{equation}
where $L_n$ represents the minimum number of bits required to encode the representation. 
We assume that the latent representations are approximately Gaussian for analytical tractability. From rate-distortion theory, we know that the minimum number of bits required to have a distortion less than $\epsilon_n$ for a Gaussian random variable with variance $\sigma^2_{z_n}$ is \cite{cover1991rate},$L_n \approx \frac{d_n}{2}\log_2\left(\frac{\sigma^2_{z_n}}{\epsilon_n}\right)$. We can now define the probability of successful transmission:  
\begin{equation}\label{reliab}
    p_n^{\textrm{succ}}(r_{n,t}) = \textrm{Pr}(\gamma_{n,t} \geq \gamma_n^{\text{req}}(r_{n,t})),
\end{equation}
where $\gamma_n^{\textrm{req}}$ is the minimum SINR required to deliver robot $n$'s features with delay $D_{\textrm{max}}$,
\begin{equation}\label{req}
    \gamma_n^{\textrm{req}}(r_{n,t}) = 2^{\frac{L_n}{r_{n,t}B_m D_{\textrm{max}}}} - 1.
\end{equation} 
Next, we introduce the IV metric that quantifies each robot's marginal contribution to inference performance, which is then used as a basis for IV-aware resource allocation.
\vspace{-4mm}
\section{Shapley Information Value metric}
\vspace{-1mm}
The BS receives the features from the robots and aims to infer $y_t$ at time step $t$, computed via the inference function $f$ on the set of all received feature vectors from all robots at the current time step. To allocate RBs to robots at each time step, the BS must determine the importance of a robot's received features relative to the features aggregated from other robots. This is because observations from nearby robots may be correlated, and allocating RBs to all of them does not efficiently utilize limited communication resources. We define the notion of importance in terms of the contribution of the feature to the application's performance. The accuracy of the inference result measures the predictiveness or value of the feature. Let $V$ be an arbitrary measure of predictiveness (e.g., accuracy or negative loss). We propose the IV metric based on the Shapley value, where the importance $\psi_{j, 0}$ of feature $j$ is defined as the average marginal gain in predictiveness across all possible subsets of features \cite{williamson2020efficientnonparametricstatisticalinference}. The IV metric  is given by,
\begin{equation}\label{shapleyeq}
    \psi_{j, 0} = \sum_{S \subseteq \mathcal{N} \backslash \{j\}} \tfrac{1}{N} \tbinom{N - 1}{|S|}^{-1} \! \left( V(f_{S \cup \{j\}}) - V(f_{S}) \right)
    \vspace{-1mm}
\end{equation}
Here, $|\mathcal{N}| = N$. Here, the function $f_{S}$ is the inference function using only the subset of features $S$. We select the Shapley value based on its axiomatic properties \cite{williamson2020efficientnonparametricstatisticalinference}. The uniqueness theorem states that the Shapley value is the only allocation that simultaneously satisfies efficiency ($\sum_n \psi_n = V(f,P))$) and symmetry (importance is independent of ordering), null player (redundant robots receive zero IV), and additivity (importance can compose across tasks). For our problem, the null player axiom correctly assigns zero IV to redundant robots that add no value to any coalition, the efficiency axiom allows the BS to distribute the total inference performance as IV to individual robots, and the symmetry axiom ensures that two robots making identical contributions to any coalition have the Shapley value regardless of ordering. 
We assume that the BS obtains an initial estimate of the importance of every robot's feature by sampling several subsets and computing the predictiveness $V$ with the inference model as,$\boldsymbol{\psi}_{0}= [\psi_{1,0},...,\psi_{N,0}]$.
\par However, direct computation of \eqref{shapleyeq} is computationally prohibitive for real-time resource allocation as the number of users increase. Moreover, the importance of individual robots may evolve over time as the wireless environment and robot positions change. The BS must update its estimates of IV in real-time.
Thus, inspired by the weighted linear regression formulation of the Shapley value in \cite{lundberg2017unifiedapproachinterpretingmodel}, we propose an online algorithm based on stochastic gradient descent (SGD) to track the Shapley value of features in real time. SGD provides a computationally simple ($O(N)$) way to track the non-stationarity of the features, which can be easily implemented in real-time. At each time step $t$, let  $\boldsymbol{k}_t \in \{0,1\}^N$ be the binary scheduling vector where each entry corresponds to a robot (1 if scheduled ($r_{n,t} > 0)$, 0 otherwise), and $V_t$ be the observed accuracy of the inference at time $t$. Then, the instantaneous squared error of the Shapley prediction model
\begin{equation}
    \mathcal{J}(\boldsymbol{\psi}) = \frac{1}{2}\cdot \pi_t (V_t - \boldsymbol{k}_t^T \boldsymbol{\psi})^2,
\end{equation}
where $\psi_{t-1}$ are the Shapley values of all features at a previous time step $\pi_t = \frac{N- 1}{\binom{N}{K} \cdot K (N - K)}$ is the Shapley weighing factor, where $K = \sum \boldsymbol{k}_t$ represents the total number of scheduled users at time $t$, and $\xi_t = V_t - \hat{V}_t$, where $\hat{V}_t = V(\emptyset) + \boldsymbol{k}_t \cdot \boldsymbol{\psi}_{t-1})$. $\xi_t$ quantifies the discrepancy between the expected and the realized performance of the scheduled set. We assume that the realized performance is available as instantaneous feedback to the remote inference model; in cases where that is not possible, $V_t$ can be approximated using proxies such as inference confidence. In our case, in the scenario where no robots are scheduled, the result is the null baseline: $V(\emptyset) = 0$. To update our estimate, we compute the gradient of the instantaneous loss with respect to the Shapley importance vector $\boldsymbol{\psi}_t$ as, $\nabla_{\psi} J(\psi) = \pi_t (V_t - \boldsymbol{k}_t^T\boldsymbol{\psi}) \boldsymbol{k}_t = - \pi_t \cdot \xi_t \cdot \boldsymbol{k}_t.$
Then the Shapley value at time $t$, $\boldsymbol{\psi}_{t} := [\psi_{1,t}, .., \psi_{N,t}]$ for all the features is given as, 
\vspace{-3mm}
\begin{equation}\label{onlineupdate}
    \boldsymbol{\psi}_t = \boldsymbol{\psi}_{t-1} + \alpha_t \cdot \boldsymbol{k}_t \cdot \pi_t \cdot \xi_t \cdot \epsilon,
\end{equation}
where $\alpha_t$ is the learning rate for tracking the update, $\epsilon$ is the chosen exploration-exploitation tradeoff parameter to update the IV of users not being scheduled. If $\boldsymbol{\psi}_0$ is initialized uniformly, $\alpha_t$ can be chosen according to a learning-rate decay schedule to encourage faster convergence. While the Shapley estimates $\psi_{n,t}$ identify which robots contribute most to inference, leveraging them into a scheduling policy is non-trivial, as both $\psi_{n,t}$ and $\gamma_{n,t}$ are time-varying, and the scheduler must account for network constraints. In the next section, we formulate the sequential stochastic IV-aware RB allocation problem as a constrained Markov decision process (CMDP).
\vspace{-8mm}
\section{CMDP Problem Formulation}
\vspace{-1.5mm}
  In our proposed approach, the BS uses a scheduling policy $\pi$ that maximizes the long-term average IV-weighted transmission success across the network subject to an RB budget and latency constraints. Formally, the scheduling objective is,
\vspace{-1mm}
\begin{subequations}\label{p1_problem}
\begin{align}
&\max_{\pi} \limsup_{T \to \infty} \tfrac{1}{T} \sum_{t=0}^{T} \mathbb{E}_{\pi} \! \left[ \sum_{n \in \mathcal{N}} \psi_{n,t}  p_n^{\textrm{succ}}(r_{n,t}) \right], \label{p1_obj} \\[-0.5em]
&\text{s.t.} \sum_{n \in \mathcal{N}} r_{n,t} \leq M, \quad \forall t, \label{p1_c1} \\[-0.5em]
&\limsup_{T \to \infty} \tfrac{1}{T} \sum_{t=0}^{T} \mathbb{E}_{\pi} \! \left[ \sum_{n \in \mathcal{N}} \psi_{n,t} T_n(\gamma_{n,t}, r_{n,t})\mathbbm{1}[r_{n,t}> 0]\right] \!\leq\! \tau, \label{p1_c2} \\[-0.6em]
&\boldsymbol{r}_t \in \mathbb{Z}_{\geq 0}^{N} \label{p1_c3}.
\end{align}
\end{subequations}
The objective weighs each robot's transmission success probability by its IV. Constraint \eqref{p1_c1} captures the per-slot RB budget, \eqref{p1_c2} refers to the long-term average-IV weighted latency constraint for the set of scheduled robots, and $\tau > 0$ represents the maximum tolerable IV-weighted latency. \eqref{p1_c2} weighs each robot's transmission latency more than that of low-IV robots. The IV-weighted latency constraint is suitable for time-sensitive inference tasks, such as 3D object detection for autonomous driving. Since \eqref{p1_problem} is a constrained stochastic optimization over a sequence of scheduling decisions under uncertainty. We formalize it as a constrained Markov decision process (CMDP) \cite{altman2021constrained}. The system state is characterized by the SINR and IV at time $t$, $\boldsymbol{s}_t = [\boldsymbol{\gamma}_t, \boldsymbol{\psi}_t] \in \mathbb{R}^{N}_{+} \times \mathbb{R}^N_{+}$. The action taken is the RB allocation vector $\boldsymbol{r}_t \in \mathcal{A} = \{\boldsymbol{r} \in \mathbb{Z}_{\geq 0}^{N} | \sum_{n} r_n \leq M\}$, where $r_n \in \boldsymbol{r}_t$ represents the number of RBs allocated to robot $n$. For convenience, let $R(\boldsymbol{s}_t, \boldsymbol{r}_t) = \sum_{n \in \mathcal{N}} \psi_{n,t} p_{n}^{\textrm{succ}}(r_{n,t})$ and $C(\boldsymbol{s}_t, \boldsymbol{r}_t) = \sum_{n \in \mathcal{N}} \psi_{n,t} T_n(\gamma_{n,t}, r_{n,t})\mathds{1}[r_{n,t}> 0]$ be the immediate reward and the cost incurred at state $\boldsymbol{s}_t$ under action $\boldsymbol{r}_t$. In this problem, we cannot characterize the joint transition state kernel $p(\boldsymbol{s}_{t+1}|\boldsymbol{s}_t, \boldsymbol{r}_t)$, as future channel realizations depend on the robot mobility and the future IV. Furthermore, the optimal policy for \eqref{p1_obj} is in general a randomized mixture of two deterministic policies obtained by solving an LP with the transition kernel \cite{altman2021constrained}. This is often computationally intractable for real-time scheduling, and thus motivates the Lyapunov drift plus penalty (LDPP) solution, which converts \eqref{p1_problem} into a sequence of tractable per-slot problems without requiring the state transition kernel.
\vspace{-5mm}
\subsection{Proposed Lyapunov Drift Plus Penalty Solution}
\vspace{-1mm}
According to the LDPP method \cite{neely2010stochastic}, the time average constraint \eqref{p1_c2} is enforced by transforming it into a queue stability constraint. We introduce a virtual queue $Z_t$, such that the stability of the queue indicates the satisfaction of the constraint \eqref{p1_c2}. Let $Z_t$ represent the \emph{virtual queue} in slot $t$, which evolves as,
\vspace{-2mm}
\begin{equation}
    Z_{t+1} = \textrm{max}\{Z_t - \tau + C(\boldsymbol{s}_t, \boldsymbol{r}_t), 0\}, Z_0 = 0,
\end{equation}
We consider a quadratic Lyapunov function $L(Z_t) = \frac{1}{2}Z^2_t$. The Lyapunov function is directly proportional to the size of the virtual queue. To stabilize the queue, we must minimize the expected change of the Lyapunov function from one time slot to the next. Let $H_t = \{\boldsymbol{s}_t, Z_t\}$ denote the system state at time $t$. The one slot conditional Lyapunov drift is defined as the expected change in the Lyapunov function given the system state $H_t$ over one time slot,
\begin{equation}\label{conditional_drift}
    \Delta_{t} = \mathbb{E}\{L(Z_{t+1}) - L(Z_t)|H_t\},
\end{equation}
where the expectation is with respect to the possibly random scheduling policy $\pi$. According to the drift plus penalty method \cite{neely2010stochastic}, to obtain a scheduling policy that solves \eqref{p1_problem}, we minimize the upper bound on the drift-plus penalty function defined as,
\begin{equation}
    \phi_t = \Delta_t - \beta \mathbb{E}[R(\boldsymbol{s}_t, \boldsymbol{r}_t)|H_t],
\end{equation}
where $\beta \geq 0$ is a parameter that adjusts the trade-off between the drift and the objective function. However, directly minimizing $\phi_t$ (equivalently $\Delta_t$) over the scheduling decisions is intractable. Instead, we derive a tractable upper bound on the drift plus penalty that only depends on quantities directly observable from the state. 
\vspace{-2mm}
\begin{proposition}\label{prop:ub}
    The upper bound for the conditional Lyapunov drift in \eqref{conditional_drift} for any feasible action $\boldsymbol{r}_t \in \mathcal{A}$ and any $\beta \geq 0$ is,
    \vspace{-1mm}
    \begin{equation}
        \Delta_t \leq B + Z_t[\mathbb{E}[(C(\boldsymbol{s}_t, \boldsymbol{r}_t) | H_t) - \tau] 
    \end{equation}
\end{proposition}
\vspace{-1mm}
where $B$ is a constant.\\
\textit{Proof .} See Appendix \ref{app:proof_ub}\\
From Proposition \ref{prop:ub}, we can bound the drift-plus penalty term as,
\vspace{-3mm}
\begin{equation}\label{upper_bound_dpp}
    \begin{split}
        \phi_t \leq B + \mathbb{E}[Z_t(C(\boldsymbol{s}_t, \boldsymbol{r}_t) - \tau) - \beta R(\boldsymbol{s}_t, \boldsymbol{r}_t)|H_t] \\
        \leq \mathbb{E}[Z_t C(\boldsymbol{s}_t, \boldsymbol{r}_t) - \beta(R(\boldsymbol{s}_t, \boldsymbol{r}_t) | H_t]
    \end{split}
\end{equation}
To minimize the upper bound in \eqref{upper_bound_dpp}, we apply opportunistic minimization to the conditional expectation, i.e \eqref{upper_bound_dpp} is minimized by observing the current state $H_t$ and choose an allocation $\boldsymbol{r}_t \in \mathcal{A}$ to minimize
\begin{equation}
     Z_t \sum_{n \in \mathcal{N}} \psi_{n,t} T_n(\gamma_{n,t}, r_{n,t}) - 
     \beta \sum_{n \in \mathcal{N}} \psi_{n,t} p^{\textrm{succ}}_n(r_{n,t}),
\end{equation}

Thus, the optimal resource allocation vector is obtained by solving the deterministic problem per slot
\begin{equation}\label{perslot}
    \boldsymbol{r}_t^{*} = \argmax_{\boldsymbol{r}_t \in \mathcal{A}}\sum_{n \in \mathcal{N}}\omega_n(\boldsymbol{r}_{n,t}),
\end{equation}
where,
\vspace{-3mm}
\begin{equation}\label{perslotsoln}
\omega_{n}(r) = 
\begin{cases} 
0, & r = 0, \\
\psi_{n,t} \left[ \beta \cdot p_{n}^{\textrm{succ}}(r) - Z_{t} \cdot T_{n}(r) \right], & r \geq 1. 
\end{cases}
\end{equation}
Thus, the proposed IV-LDPP scheduler works by observing $H_t$ at every time slot $t$ and solving the deterministic per-slot problem \eqref{perslot}. Now, we will show that the solution in \eqref{perslotsoln} satisfies the constraint \eqref{p1_c2}.
\vspace{-1mm}
\begin{theorem}\label{stability_proof}
    Let $\boldsymbol{r}^{*}_t$ be the scheduling decision of solving the per slot problem at time t in \eqref{perslot}\footnote{The Slater condition holds for \eqref{p1_problem} as there exists a trivial policy such that \eqref{p1_c1}, \eqref{p1_c2}, \eqref{p1_c3} are satisfied \cite{altman2021constrained}}. Assuming that $\mathbb{E}[L(Z_0)]$ is finite. Then, for any finite $\beta \geq 0$, the virtual queue is strongly stable.
    \begin{equation}
        \limsup_{T \rightarrow \infty} \frac{1}{T}\sum_{t=0}^{T-1}\mathbb{E}[Z_t] \leq \frac{B + \beta R_{\textrm{max}}}{\tau},
    \end{equation}
    where $B = \frac{1}{2}(\tau^2 + C^2_{\textrm{max}})$ and $R_{\textrm{max}} = \sum_n \psi_{n} = V_{\textrm{max}}$, where $V_{\textrm{max}}$ is the maximum possible inference accuracy. $R_{\textrm{max}}$ denotes the maximum achievable reward per slot. Strong stability of the queue implies that the constraint $\eqref{p1_c2}$ is satisfied.
\end{theorem}
\vspace{-3mm}
\textit{Proof. } See Appendix \ref{app:proof_stability}.\\
Theorem \ref{stability_proof} establishes that the deterministic per-slot solution \eqref{perslotsoln} satisfies the constraint \eqref{p1_c2} in the long run without requiring knowledge of future states or the transition dynamics and is thus a valid solution to the objective \eqref{p1_problem}.
\vspace{-4.5mm}
\section{Simulation Results and Analysis}
\vspace{-1mm}
We evaluate the IV-LDPP scheduler in a semantic remote inference scenario. Multiple camera/LiDAR-equipped robots transmit observations to a centralized BS, which performs cooperative 3D object detection. Experiments are conducted using the nuScenes v1.0-mini dataset, which comprises RGB camera, LiDAR, and radar data \cite{nuscenes2019} across 10 scenes, each 20s long, along with annotations for bounding boxes for different object classes (e.g., pedestrian, car, truck). At the BS, we deploy the pretrained BEVFusion model \cite{liu2023bevfusion}, which fuses camera and LiDAR features in bird's eye view (BEV) space. We consider a set of $|\mathcal{N}| = 7$ robots that sample RGB images and LiDAR data from the 6 distinct cameras and one LiDAR sensor, respectively. We consider an uplink channel with numerology $\mu = 1$ (30 kHz subcarrier spacing). Each RB occupies $B_m = 12 \times 30 \text{kHz} = 360 \text{kHz}$. The total RB budget is $M = 5$, which is shared among $N = 7$ robots. We evaluate the scheduler over an SNR sweep from $-5$ dB to 20 dB to capture varying channel conditions across $T=81$ time slots \footnote{nuScenes mini validation set has 81 unique samples \cite{nuscenes2019}.}. We compare the IV-LDPP scheduler against the following baselines, 
\vspace{-1mm}
\begin{itemize}
    \item \emph{MaxRate}: Schedules robots with the best instantaneous channels.
    \item \emph{AoI}: Prioritizes the least fresh robot.
    \item \emph{TAoI}: AoI counter resets only when downstream inference improves \cite{10901473}
    \item \emph{VoIU}: Uses the marginal improvement in accuracy for scheduling \cite{9068225}. 
\end{itemize}
We evaluate the scheduling performance using the mean average precision (mAP) score, which quantifies the overlap between predicted and ground-truth boxes. We simulate two variants of our proposed IV-LDPP scheduler: Shapley prior, using Shapley value precomputed offline on the dataset, and Uniform, with the equal Shapley value across all robots and updates via \eqref{onlineupdate}. The offline Shapley computation shows that LiDAR and front camera sensors are more important than the rest. Specifically, the LiDAR sensor has significantly higher IV than the camera sensors, $\psi_{\textrm{LiDAR}} >> \psi_{\textrm{cam}}$, which follows the fact that LiDAR is more informative than camera for 3D object detection.
\begin{figure}[t]
    \centering
    \includegraphics[width=0.48\textwidth]{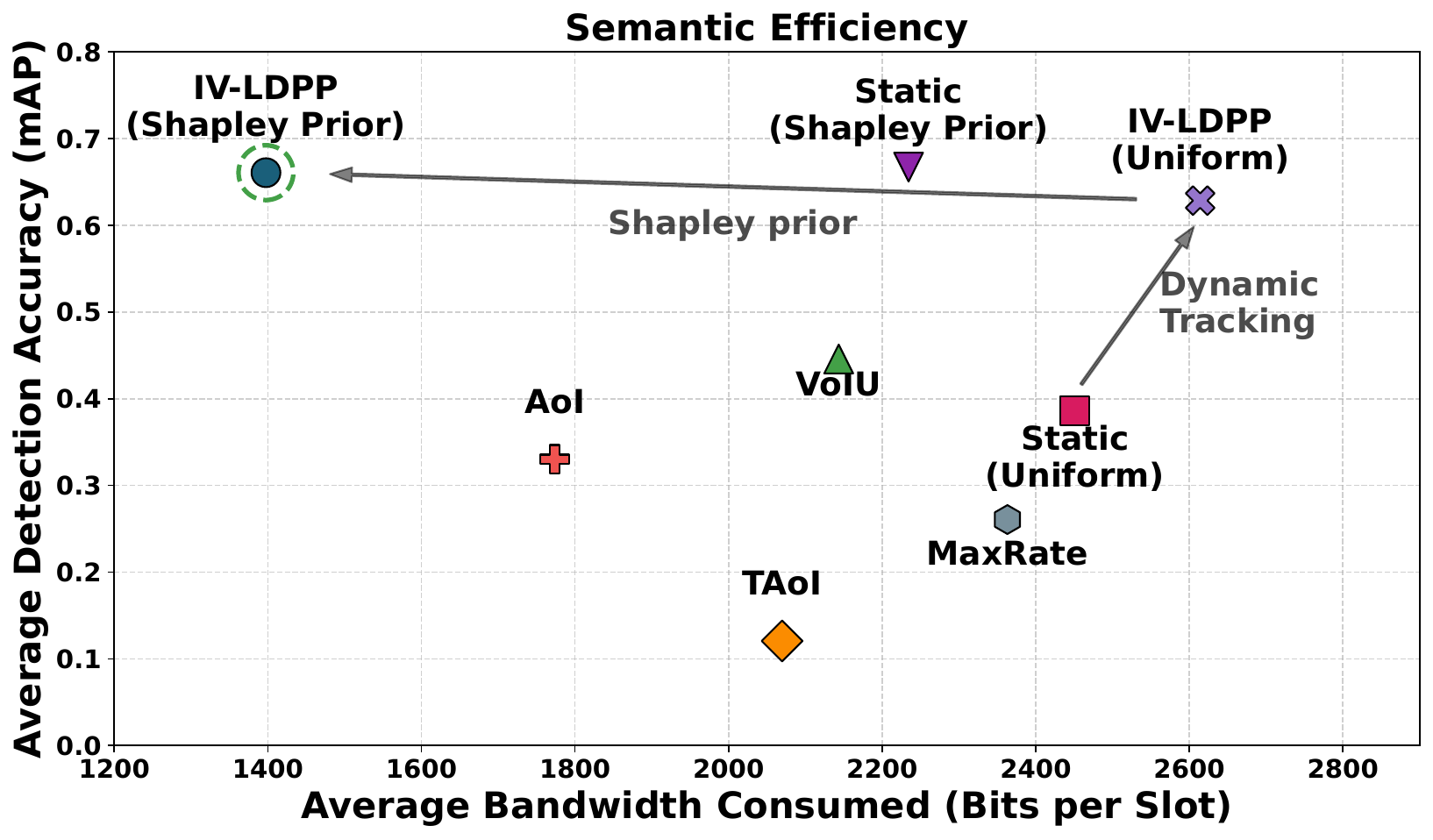}
    \caption{Scatter plot showing detection accuracy vs. bits consumed across all baselines.}
    \label{fig:mapvsbandwidth}
    \setlength{\belowcaptionskip}{-15mm}
    \vspace{-3mm}
\end{figure}
\par Fig. \ref{fig:mapvsbandwidth} compares the bandwidth consumption against mAP averaged across different SNRs. Our proposed IV-LDPP (Shapley prior) algorithm achieves a 53\% higher mAP score than the VoIU scheduler while using 43.2\% less bandwidth compared to the MaxRate scheduler. AoI scheduler uses the least bandwidth but achieves a low mAP score, confirming that transmission freshness alone is insufficient for better inference. The MaxRate scheduler utilizes the most bandwidth; however, it achieves a lower mAP score than our proposed IV-LDPP scheduler, since it allocates RBs by channel quality and not feature importance. We can also see that the Shapley online update from \eqref{onlineupdate} helps the scheduler converge to only the most important robots for inference, as scheduling with static Shapley value performs similarly but is bandwidth inefficient. Moreover, for the IV-LDPP (Uniform) scheduler, the update in \eqref{onlineupdate} helps achieve better performance while starting with a non-informative point. Initially, the IV-LDPP (Uniform) scheduler uses more bandwidth because it treats all robots as equally important. As the online update \eqref{onlineupdate} progressively concentrates importance on the LiDAR stream over $T=500$ slots, IV-LDPP (Uniform) eventually converges to the same operating point as IV-LDPP (Shapley prior), as shown by the arrow in Fig. \ref{fig:mapvsbandwidth}.
\par
\begin{figure}[!t]
    \centering
    \includegraphics[width = 0.48\textwidth]{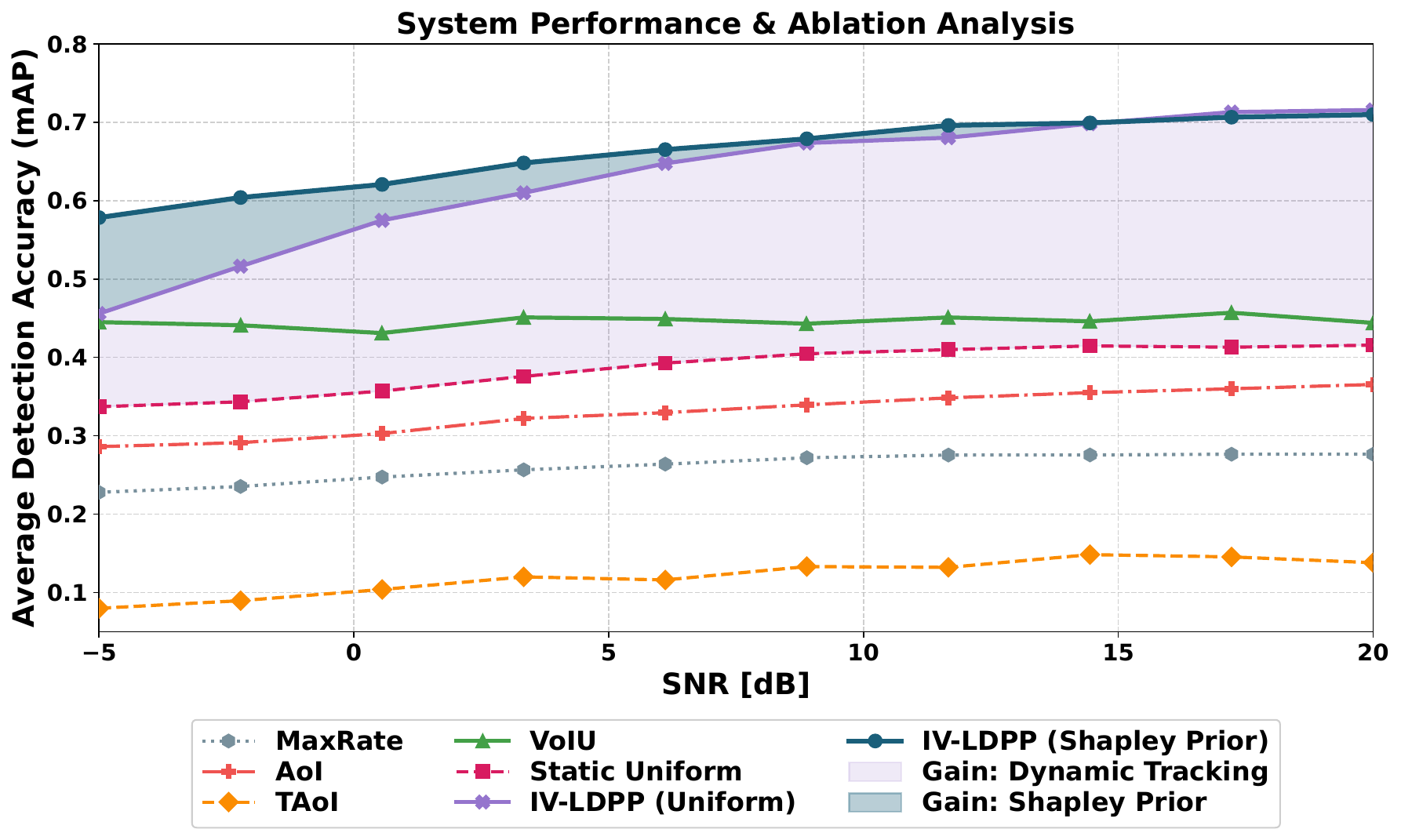}
    \caption{Detection accuracy vs. SNR comparison across all baselines.}
    \setlength{\belowcaptionskip}{-15mm}
    \label{fig:mapvssnr}
    \vspace{-4mm}
\end{figure}

 Fig. \ref{fig:mapvssnr} shows the downstream task performance across varying channel conditions ranging from an SNR of -5 dB to 20 dB. The proposed IV-LDPP scheduler outperforms all the baselines across all SNRs. This is achieved because the IV-LDPP scheduler can identify that the LiDAR feature stream is more critical for the downstream task than the other camera data streams, as it has a higher IV than the camera data streams. From Fig. \ref{fig:mapvssnr}, we see that the update in \eqref{onlineupdate}, helps achieve a better detection accuracy for the IV-LDPP (uniform) scheduler. The difference in task performance between IV-LDPP (Shapley prior) and IV-LDPP (Uniform) is only significant at lower SNRs, and the gap reduces to being negligible in good channel conditions. However, the IV-LDPP (Uniform) scheduler may be bandwidth inefficient if it has not fully converged. 
\par
\begin{figure}[!t]
    \centering
    \includegraphics[width=0.48\textwidth]{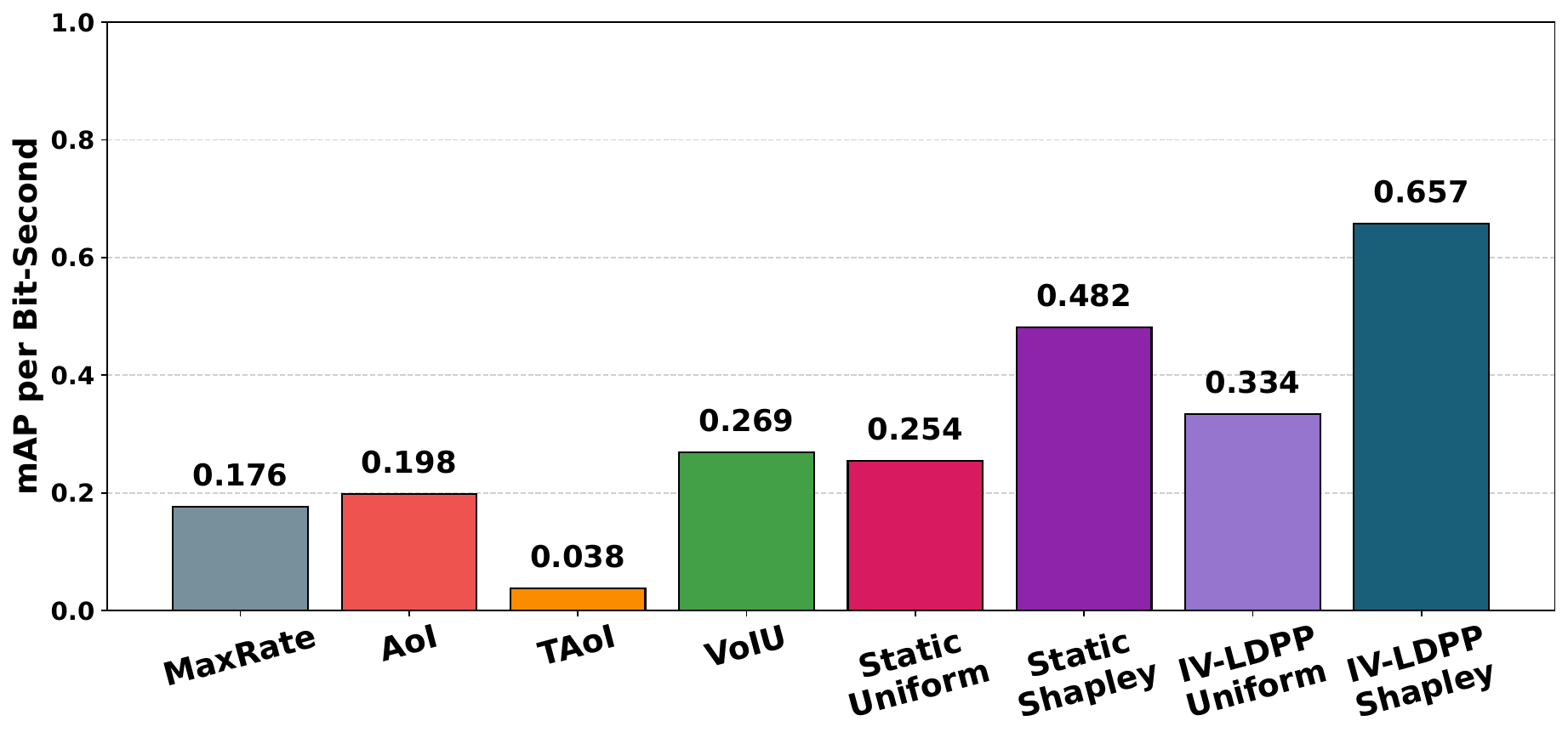}
    \vspace{-3mm}
    \caption{Network efficiency (mAP per bits-second) comparison across all baselines, measuring the inference gain per unit of consumed bandwidth and latency.}
    \setlength{\belowcaptionskip}{-11mm}
    \label{networkefficiency}
    \vspace{-4mm}
\end{figure}

 Fig. \ref{networkefficiency} shows the task performance per bits per latency incurred. Fig. \ref{networkefficiency} demonstrates that the IV-LDPP achieves a better network efficiency than MaxRate and AoI. The proposed IV-LDPP (Shapley prior) scheduler achieves the best efficiency among all schedulers, confirming that IV-aware scheduling delivers more inference gain per consumed resource than channel quality or freshness-based alternatives. 
%
\vspace{-5mm}
\section{Conclusion}
\vspace{-2mm}
In this paper, we have developed an IV-aware scheduler for task-oriented remote inference systems. We have proposed an IV metric based on the Shapley value to quantify feature importance and utilize it for resource allocation. We have posed the IV-aware resource allocation problem as a CMDP and solve it using the Lyapunov drift plus penalty method. Simulations for 3D object detection task using the BEVFusion inference model and the nuScenes dataset show that the proposed IV-LDPP scheduling algorithm outperforms the traditional capacity and freshness baselines in downstream task performance while using significantly fewer resources, as it can exploit the IV of the transmitting agents to only schedule agents that contribute the most to the downstream task. 

\vspace{-4mm}
\appendices
\vspace{-2mm}
    \section{Proof of Proposition \ref{prop:ub}}\label{app:proof_ub}
    \vspace{-2mm}
    \begin{proof}
        To derive the upper bound for $\Delta_t$, we use the inequality in \cite{neely2010stochastic}, for any $A  \geq 0$, $B \geq 0$, $C \geq 0$, we have, $ (\textrm{max}\{A - B + C, 0\})^2 \leq A^2 + B^2 + C^2 + 2A(C - B)$. By applying this,to eq. \eqref{conditional_drift},
        \begin{equation}
    \begin{split}
        Z^2_{t+1} &\leq Z^2_{t} + \tau^2 + C^2(\boldsymbol{s}_t, \boldsymbol{r}_t) + 2Z_t(C(\boldsymbol{s}_t, \boldsymbol{r}_t) - \tau) \\
        &\leq B + Z_t[\mathbb{E}(C(\boldsymbol{s}_t, \boldsymbol{r}_t)|H_t) - \tau]
    \end{split}
\end{equation}
where, $B = \frac{1}{2}(\tau^2 + C^2_{\textrm{max}})$, where $C_{\textrm{max}} = \sum_{n \in \mathcal{N}} \frac{V_{\textrm{max}}L_n}{B_m \log_2 ( 1+ \gamma_{\text{min}})} = V_{\textrm{max}}D_{\textrm{max}}$ is the maximum possible latency cost in the system.
    \end{proof}
    \vspace{-7mm}
    \section{Proof of Theorem \ref{stability_proof}}\label{app:proof_stability}
\begin{proof}
\vspace{-2mm}
To show the strong stability of the virtual queue, we first define the idle policy $\pi^{\text{idl}}$ such that $\boldsymbol{r}^{\text{idl}}_t = 0, \forall t$. Under this policy, $\mathcal{S}_t = \emptyset$, $C(\boldsymbol{s}_t, \boldsymbol{r}^{\text{idl}}_t) = 0$, and $R(\boldsymbol{s}_t, \boldsymbol{r}^{\text{idl}}_t) = 0$. From Proposition \ref{prop:ub}, applying the upper bound to the idle policy and noting that the optimal policy $\pi^*$ achieves a lower drift-plus-penalty value, we obtain $\Delta_t \leq B - \tau Z_t + \beta R_{\text{max}}$. Taking the unconditional expectation via the law of iterated expectations, summing over $t = 0, \dots, T-1$, and utilizing the telescoping property of $\mathbb{E}[L(Z_t)]$, we have $\tau \sum_{t = 0}^{T - 1} \mathbb{E}[Z_{t}] \leq T(B + \beta R_{\text{max}}) - \mathbb{E}[L(Z_{T})] + \mathbb{E}[L(Z_0)]$. Dividing by $T\tau$, noting $\mathbb{E}[L(Z_T)] \geq 0$, and taking the $\limsup_{T \to \infty}$ yields:
\begin{equation}
    \limsup_{T \to \infty} \frac{1}{T} \sum_{t=0}^{T-1} \mathbb{E}[Z_t] \leq \frac{B + \beta R_{\text{max}}}{\tau}.
\end{equation}
This confirms the virtual queue is strongly stable.
\vspace{-7mm}

\end{proof}
\begin{small}
    \bibliographystyle{IEEEtran}
\def\baselinestretch{0.87}
\bibliography{semantic_refs}
\end{small}





\vfill

\end{document}